# AGI, Governments, and Free Societies

Justin B. Bullock[1], Samuel Hammond[2], and Seb Krier

## Abstract

This paper examines how artificial general intelligence (AGI) could fundamentally reshape the delicate balance between state capacity and individual liberty that sustains free societies. Building on Acemoglu and Robinson's 'narrow corridor' framework, we argue that AGI poses distinct risks of pushing societies toward either a 'despotic Leviathan' through enhanced state surveillance and control, or an 'absent Leviathan' through the erosion of state legitimacy relative to AGI-empowered non-state actors. Drawing on public administration theory and recent advances in AI capabilities, we analyze how these dynamics could unfold through three key channels: the automation of discretionary decision-making within agencies, the evolution of bureaucratic structures toward system-level architectures, and the transformation of democratic feedback mechanisms.

Our analysis reveals specific failure modes that could destabilize liberal institutions. Enhanced state capacity through AGI could enable unprecedented surveillance and control, potentially entrenching authoritarian practices. Conversely, rapid diffusion of AGI capabilities to non-state actors could undermine state legitimacy and governability. We examine how these risks manifest differently at the micro level of individual bureaucratic decisions, the meso level of organizational structure, and the macro level of democratic processes.

To preserve the narrow corridor of liberty, we propose a governance framework emphasizing robust technical safeguards, hybrid institutional designs that maintain meaningful human oversight, and adaptive regulatory mechanisms. Our recommendations focus on practical steps to ensure AGI deployment strengthens rather than subverts democratic values, from privacy-enhancing technologies to new models of citizen participation. We conclude that maintaining free societies in an age of AGI requires deliberate institutional innovation to harness its benefits while guarding against both centralized control and institutional collapse.

[1] Texas A&M University; Convergence Analysis; Global Governance Institute
[2] Foundation for American Innovation

# I. Introduction

Throughout history, leaps in technology have destabilized and altered both the politics and administration of governments. Many current observers expect artificial intelligence to do the same. However, the present conversation on AI governance has focused on how current AI models may be utilized by governments; few have explored the ways in which artificial general intelligence (AGI) that meets or exceeds the capabilities of humans on all decision making tasks may lead to further evolutions in both government politics and the administration. Even less work examines how AGI might impact the narrow corridor that leads to free and open societies in the liberal tradition. We seek to remedy this gap by exploring the plausible consequences to governments in a world with AGI and other powerful forms of AI.

Governments are composed of two key, but indistinct, aspects: politics and administration (Wilson, 1887). Politics is defined by who decides, who makes the rules, and how these individuals are selected. Administration refers to choices, actions, and institutions. Over time, these concepts have evolved to create governmental forms, from tribes and divine monarchies to republics, constitutional monarchies, and liberal democracies.

A recent emergence in the history of human civilization, the liberal political tradition constrained governments to respect individual political and economic rights and liberties amid the 17th and 18th century acceleration in science and technology, giving rise to societies that were both freer and more prosperous than what came before. The parallel demands for efficient public administration and political participation were thus secured through a 'narrow corridor' that reconciled the relative powers of society and the state, with concepts such as the rule of law and limited government representing a balancing act between anarchy and authoritarianism (Acemoglu & Robinson, 2019). As a path-dependent and technological-contingent equilibrium, there is no guarantee that liberal institutions will survive over the long-term. Future developments have the potential to reshape the balance of power in the direction of either a 'despotic Leviathan' or an 'absent Leviathan.'

One such force shaping the balance of power is technological change, which began gaining speed around the same time as liberalism emerged as a political philosophy. Key technological innovations such as the steam engine, electric light, and the telephone began altering both the politics and administration of governments in the 1800s. By the mid-1900s, electronic computing machines and the nuclear bomb were developed, forever altering both governments and societies (Roser, 2023).

In what follows, we describe the plausible near-term capabilities of AI, and explore what sorts of capacities an AGI is likely to have. We then consider the key features of open and free societies over the past 500 years, including their mechanisms of development and maintenance. With these mechanics in mind, we then examine how AGI and similarly powerful forms of narrow AI might impact both the politics and administration of governments. Finally, we provide recommendations to both governments and AI developers, development and deployment strategies that may strengthen and enhance freedom, openness, and the liberal tradition, instead of destroying them.

## II. The Age of Artificial Intelligence

The field of artificial intelligence emerged in the mid-1900s, with the primary goal of approximating the decision making capabilities of humans, an ability often described as artificial general intelligence or AGI. The second half of the 20th century and the first decade of the 21st saw incremental, if uneven, progress towards this goal. Even so, notwithstanding several significant breakthroughs, models remained limited in capabilities and narrow in their decision making scope.

However, the late 2010s brought dramatic increases in both capabilities and generality, in the form of large language models (LLMs) built using generative pre-trained transformers (GPTs). This leap forward has led many to claim that "sparks" of AGI have combusted that will soon achieve AGI, or even superintelligent capabilities. That is, we already have AI models that can approximate and even surpass the decision making capabilities of humans in many domains, and seemingly sit on the cusp of AI systems that can do this in *every* decision making domain.

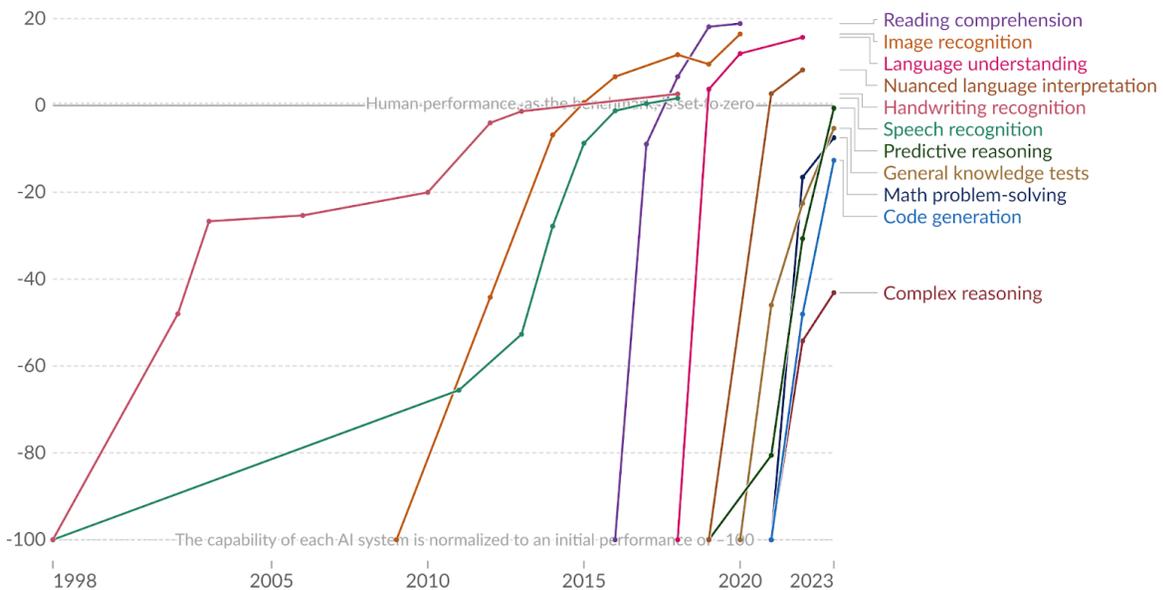

AI systems have already taken over many roles, tasks, and functions once performed exclusively by humans, and are integrating rapidly into complex industries. In air travel, they determine ticket prices, monitor passengers at airports, and assist pilots in flight. Though these systems arguably still lack the general intelligence and adaptability that humans possess across varied situations, as the above illustration shows, the gaps are rapidly narrowing. Recent interest in agentic architectures is also showing some early promise: Anthropic's Claude 3.5 Sonnet model, for example, achieved 49% on SWE-bench Verified, an evaluation targeting real-world software engineering tasks.

Morris et al.[3] define AGI through a multi-dimensional framework that considers both performance and generality. For the purposes of this paper, we are interested in the 'Expert AGI' level: that is, a system that demonstrates performance at or above the 90$^{th}$ percentile of skilled adults across a wide range of cognitive tasks, representing a significant milestone in AGI development. But how far are we from such capabilities? Rapid improvements in models have surpassed the expectations of many observers and academics in the field, largely due to the sheer scale of the data and computing resources available to train deep neural network models.

Scaling *laws*, in their formal sense, are mathematical relationships that predict how a model's performance (often measured by test loss) changes with factors like model size, dataset size, and training compute. Notably, these laws have exhibited a consistent pattern across various domains, including language modeling, computer vision, and reinforcement learning—demonstrating that power laws (albeit with different coefficients and exponents) exist in these domains. More generally however, *'scaling'* refers to the approach of improving AI capabilities by increasing the scale of inputs within existing model paradigms. This includes increasing model size, data, and compute, and investing in post-training techniques like prompting, tool use, and scaffolding (i.e., 'unhobbling' the model). Crucially, recent research has expanded the understanding of scaling to include the strategic allocation of compute during inference. This 'test-time compute scaling' involves techniques that allow models to utilize additional computational resources when processing a specific prompt, leading to improved performance. Snell et al. (2024) demonstrate that, in certain cases, optimizing inference-time compute can be more effective than simply scaling model parameters.[4]

In short, bigger is better—at least so far—thanks to a combination of technological advancements and increased resource allocation, leading to rapid progress in AI capabilities. The compute required for language models to reach a set performance threshold has halved approximately every eight months since 2012. In addition to significant algorithmic innovations, such as the Transformer architecture, many recent performance gains stem primarily from the ability to scale up models and datasets to unprecedented sizes. This has been made possible by rapid growth in computing power, which has allowed the rate of improvement to substantially exceed hardware gains from Moore's Law.[5]

---

[3] https://arxiv.org/abs/2311.02462
[4] https://arxiv.org/abs/2408.03314v1

Countless recent advances by labs, academic institutions, and open source communities reflect growing investment in the field and a diversification of approaches and algorithms. For example, new systems from Google DeepMind have recently solved four out of six problems from last year's International Mathematical Olympiad, matching the competition's silver medalist for the first time.[6] Yet in this case, the most significant breakthroughs often come from clever combinations of techniques, domain-specific optimizations, and novel approaches to bridging different AI paradigms, rather than scaling *per se*.

The effectiveness of post-training enhancements highlights that raw training compute is not the sole determinant of model capabilities. Clever prompting, tool integration, and other techniques can significantly amplify performance at a fraction of the training cost. Davidson and colleagues analyze various 'post-training enhancements' across five categories: tool use, prompting methods, scaffolding, solution selection, and data generation.[7] Most of the enhancements studied produced improvements equivalent to increasing the original training compute by a factor of five, and often by much more. These enhancements often cost less than 1% of the original training compute, making them highly efficient ways to boost model capabilities in specific domains.

As noted above, improvements in model performance can be achieved not only through increased training compute, but also through more efficient use of inference time compute, the cost of which is many times cheaper than training. The remarkable performance of models like o1, Gemini 2.0 Flash Thinking, and o3 can be partially attributed to these inference-time optimizations, suggesting that there is still substantial room for improvement in this domain without necessitating the enormous costs associated with increased training compute.

## How far will these scaling laws take us?

The potential of scaling as a general approach is more complex. While the rapid progress of recent years suggests significant headroom, potential bottlenecks include data availability, the development of fundamentally new architectures, and unforeseen limitations in current paradigms. The role of factors beyond compute, such as innovative training techniques and improved understanding of intelligence itself, becomes increasingly important. There are good reasons to think that future improvements may stem from identifying and developing better and inherently scalable architectures rather than increasing the size of existing models indefinitely.[8] In addition,

---

it is difficult to reliably predict performance on specific downstream tasks (i.e., translating 'test loss' or perplexity into a model's capability for analyzing a paper or writing poetry).[9]

On the other hand, recent research suggests that such capability predictions could be more feasible than previously thought. Epoch investigated language model performance, across five orders of magnitude of compute scaling, in 11 recent model architectures, and found that aggregate benchmark performance is fairly predictable, with mean absolute errors of around six percentage points when predicting across an order of magnitude of scaling. Even individual benchmark tasks, while more variable, were found to be more predictable than chance. This suggests that while challenges remain in predicting *specific* capabilities, careful extrapolations based on benchmark performance can still provide valuable insights into future AI performance.[10] For example, Hernandez et al. (2021) study scaling laws in the context of transfer learning between language tasks.[11] They found that the amount of data effectively transferred from the source task to the target task scales as a power law with respect to both fine-tuning data and model size. This suggests that leveraging scaling laws can lead to more efficient and effective transfer learning, potentially improving language model capabilities across tasks.

Progress in AI capabilities over the past few years is starkly illustrated by performance improvements on most standardized benchmarks. For instance, on the MATH benchmark testing competition-level mathematics, model performance skyrocketed from a mere 6.9% in 2021 to 84.3% in 2023. Similarly dramatic improvements were seen in other domains: grade school math problem-solving (GSM8K) jumped from 66.6% to 97% in just one year, while code generation abilities (HumanEval) soared from 32.2% to 96.3% between 2021 and 2023. Perhaps most impressively, on the MMLU benchmark assessing broad knowledge and reasoning, AI surpassed the human baseline in 2023, with scores rising from 32.44% in 2019 to 90.04%. These leaps in performance underscore the swift and substantial advancements in AI capabilities.[12]

While some dispute whether the model performance on these benchmarks is attributable to genuine reasoning or memorization, recent evidence shows that models exhibit a complex interplay between these two capabilities. Indeed, they are not necessarily opposing forces—models appear to leverage both mechanisms, with memorization potentially helping bootstrap the development of more general reasoning capabilities.[13] Further, performance on benchmarks that require multi-step, open-ended reasoning suggests that models are developing more sophisticated problem-solving abilities beyond simple memorization.

---

[9] https://arxiv.org/abs/2406.04391
[10] https://epochai.org/blog/ai-capabilities-can-be-significantly-improved-without-expensive-retraining (https://arxiv.org/abs/2312.07413)
[11] https://arxiv.org/html/2407.02118v1
[12] https://aiindex.stanford.edu/report/#individual-chapters
[13] https://arxiv.org/pdf/2410.23123

A blog post by Dwarkesh Patel neatly outlines several cruxes in the debate over how far scaling can take us.[14] The question around the future of AI hinges on several interconnected factors, going beyond simply scaling compute. Data availability and quality are crucial, as are advancements in areas like self-supervised learning, reinforcement learning, and more efficient and robust architectures. Furthermore, societal factors like regulation, ethical considerations, and public acceptance will play a significant role in shaping the trajectory of AI development. Analysis by Epoch examines four potential blockers to AGI scaling: power constraints, chip manufacturing capacity, data scarcity, and the latency wall. Their findings suggest that training runs of up to 2e29 FLOP will likely be feasible by 2030, representing a 10,000-fold scale-up from current models. While power supply and chip manufacturing emerge as the most immediate constraints, they conclude that there is still significant scope for further training and scaling. They also note that these scales could enable AI advances by decade's end as substantial as those seen since its beginning, potentially attracting hundreds of billions in investment.

Ultimately, there remain important uncertainties which, combined with questions about hardware limitations, make it challenging to confidently predict the ultimate limits of the scaling approach. More fundamentally, the very nature of intelligence itself is contested, with some viewing it as emergent from efficient data compression, while others argue it requires additional, possibly yet-undiscovered mechanisms. So far, however, there are few signs of a slowdown, and compelling reasons to believe that progress towards AGI is unlikely to slow materially.

## Real-world impacts: where's my flying car?

Importantly, impressive benchmarks and capabilities do not necessarily translate into immediate, large-scale industrial applications or societal impacts. Amara's Law aptly states that we tend to overestimate the effect of a technology in the short run and underestimate it in the long run. This is particularly relevant when considering the transition from 'useful and impressive new discovery' to 'scalable and profitable production tool and manufacturing process.' In the short term, this process is often protracted, due to various factors such as adoption rates, user experience refinement, cultural adjustments, regulatory barriers, and the need to develop supporting infrastructure and pipelines. As such, there is typically a lag between the demonstration of new AI capabilities and their widespread, value-adding implementation in industry and society.

Other factors can further impede the translation of benchmark results to real-world applications.[15] First, many AI models excel in controlled environments but struggle with the variability and unpredictability of real-world scenarios. This 'reality gap' often requires significant additional development to bridge. Second, the integration of AI systems into existing business processes and legacy systems is often complex and time-consuming, requiring substantial investment and organizational change and constrained by a shortage of skilled professionals who can effectively implement and manage these advanced AI systems in practical settings. In practice, while workers see substantial productivity potential in ChatGPT, adoption is often hindered by employer restrictions and training needs.[16] A recent field evaluation of the government integration of a fraud detection system showed that even when a model correctly identified fraudulent firms and inspections confirmed their non-existence, there were delays and inconsistencies in the tax department's follow-up actions: they had no clearly articulated processes for tracking and acting on these flags in the first place.[17] Third, public trust and acceptance of AI technologies can be slow to build, particularly in sensitive areas like healthcare or finance.

---

[14] https://www.dwarkeshpatel.com/p/will-scaling-work
[15] https://mattsclancy.substack.com/p/when-the-robots-take-your-job

These factors collectively contribute to a considerable lag between benchmark achievements and tangible societal impact, explaining why despite rapid progress in AI capabilities, we might not yet see dramatic changes in our day-to-day lives. Even so, a growing number of studies observe important 'real-life' effects, such as significant productivity increases for coders.[18] Further, this lag does not negate the potential for accelerating progress in the medium to long term. In fact, in many ways, AI's impact to date remains underestimated. First, it is not easy to gather data sufficiently granular to understand how basic language models are changing businesses and employee work. Much empirical work is still needed to understand impacts on labor markets.[19] Secondly, media attention on consumer-oriented use cases and chatbots misses the many ways AI systems are used in science,[20] humanities, history,[21] and engineering.[22] In economic research,[23] AI agents are being used to simulate[24] complex market dynamics and test policy interventions. In cybersecurity, an AI agent has, for the first time, discovered a previously unknown, zero-day, exploitable memory-safety vulnerability in widely used real-world software.[25] And in fields like drug discovery, AI systems are already screening millions of potential compounds to identify promising candidates much faster than traditional methods.[26]

A recent field experiment with over 750 management consultants found that AI assistance increased task completion rates by 12.2% on average and improved output quality by over 40% compared to a control group. However, the study also revealed that AI can decrease performance on tasks outside its current capabilities, highlighting the importance of carefully navigating AI's

---

[16] https://papers.ssrn.com/sol3/papers.cfm?abstract_id=4827166
[17] https://www.nber.org/papers/w32705
[18] https://papers.ssrn.com/sol3/papers.cfm?abstract_id=4945566
[19] https://www.governance.ai/post/predicting-ais-impact-on-work
[20] https://royalsociety.org/-/media/policy/projects/science-in-the-age-of-ai/science-in-the-age-of-ai-report.pdf
[21] https://scrollprize.org/
[22] https://arxiv.org/abs/2403.17927
[23] https://www.nber.org/system/files/working_papers/w30957/w30957.pdf
[24] https://www.nber.org/papers/w32381
[25] https://googleprojectzero.blogspot.com/2024/10/from-naptime-to-big-sleep.html
[26] https://pmc.ncbi.nlm.nih.gov/articles/PMC10302890/

'jagged frontier' of abilities.[27] While these applications may not show up directly in GDP figures *yet*, they are laying the groundwork for major advances and productivity gains in the coming years.

In our view, these large-scale impacts are more likely than not: it seems reasonable to expect continued progress across many fronts (e.g., pre-training, data curation, post-training enhancements, deployment, and adoption) and various dimensions (e.g., improved architectures, compute and training efficiencies, integration in industrial processes, etc.). As Sabine Hossenfelder notes: "AGI or not, any system capable of logical reasoning in verbal form, identifying logical fallacies and biases will change the world dramatically & very quickly."[28] So what should we expect to see over the coming years? How close are we to AGI? How should public administration and management scholars consider the implications? We explore these questions below.

## Next stop: AGI

Understanding AGI requires moving beyond a binary conception to a more incremental framework that considers both the breadth (generality) and depth (performance) of AI capabilities. As discussed above, Morris and colleagues proposed six levels of AI performance (No AI/Emerging/Competent/Expert/Virtuoso/Superhuman) based on key principles: focusing on *capabilities* rather than *processes*, considering cognitive and metacognitive abilities, evaluating *potential* rather than *deployment*, and using ecologically valid benchmarks. A common trap in AI governance discussions is to conflate capabilities with deployments, and deployments with economic impacts. Definitions aside, what does this mean for politics, governance, and public administration?

We suggest that as AI systems advance through these levels, acquiring increasingly broad and sophisticated capabilities, the very nature of governance and decision-making is likely to be transformed as well. This technology could enable more creative approaches to problem-solving, allowing for the assignment of AI agents and systems to tasks and functions that previously relied on slower, less efficient human processes. In principle, a manager may have at their disposal what is effectively a much larger supply of 'cognitive labour' to apply to a wide array of problems. Importantly, the ability of models to improve their outputs by using more test-time computation suggests a pathway towards achieving deeper and more reliable performance in complex tasks, bringing us closer to more advanced levels of AGI.

This shift has the potential to transform how governments operate, analyze complex issues, and implement policies, for better or for worse. Governments grappling with AI policy should therefore think beyond regulation, embracing a creative 'futurist' mindset that anticipates near-AGI capabilities within the next decade. This involves reimagining governance and policy prescriptions in light of AI agents potentially undertaking many (if not most) screen-based tasks currently performed by humans. We suggest AGI will affect governance and public administration in at least three significant ways:

1. **Deep integration *within* government decision-making**. Adopting AGI systems in daily operations will be challenging. It will require not only tailoring products to suit civil servants' workflows but also fostering a culture that embraces experimentation and reliance on these systems. In addition, governments will increasingly be responsible for maintaining the complex network of infrastructure, sensors, and systems that keep societies functioning. This includes not only physical management but also governance, oversight, and regulation, areas where AGI agents could potentially provide substantial assistance.

2. **Restructuring the machinery of government**. The very composition of government may evolve, with some departments expanding while others become obsolete. The potential for AGI to process and synthesize enormous amounts of data and perform a large number of sophisticated tasks, while accounting for multifaceted preferences and constraints, opens up possibilities for governance mechanisms that are more responsive, equitable, and effective than current models allow.

3. **Reinforcing democratic feedback loops**. The current model of representative democracy, in which, for example, a single MP acts as an agent for constituents' views in parliament, could be improved. AGI offers an opportunity to rethink and potentially strengthen both individual agency and our often ossified democratic decision-making systems, reimagining the feedback loop between citizens and their representatives. This could help address long-standing challenges in democratic systems such as voter apathy, information asymmetry, and the disconnect between elected officials and their constituents.

---

[27] https://www.hbs.edu/ris/Publication%20Files/24-013_d9b45b68-9e74-42d6-a1c6-c72fb70c7282.pdf
[28] https://x.com/skdh/status/1822562723235864984?prefetchTimestamp=1739216854490&mx=2

Whether these changes end up net positive or negative will depend on who uses the technology, and how they use it. AGI offers the potential to significantly empower both authoritarian and democratic states; part of the challenge will be ensuring the latter stays ahead. To that end, it is worth revisiting the emergence of liberalism in its historical context, as understanding how technology shaped our contemporary institutions can shed light on how the mechanisms of a free society may change going forward.

# III. Mechanics of Free Societies

## Historical pathways

The development of free and liberal democratic societies is a relatively recent phenomenon in human history. While ideas of individual liberty and constraints on power can be traced back to ancient civilizations, the modern conception of liberal democracy emerged primarily in the early modern period, the 16$^{th}$ and 17$^{th}$ centuries, with the rise of nation-states in Europe. The diffusion of the printing press and a growing mercantile class enabled administrative record keeping amid demands for property and contract enforcement, laying the groundwork for more centralized forms of governance (Dittmar, 2011).[29] Paradoxically, in the aftermath of the European wars of religion, this centralization created the conditions for constitutional constraints on power and thus the development of the modern Westphalian nation-state.[30,31] The Glorious Revolution of 1688, for instance, established the principle of parliamentary sovereignty, placed limits on monarchical authority, and enforced Acts of Toleration to stem religious conflict.[32]

In the 18$^{th}$ century, the Enlightenment catalyzed liberal thought, with a particular emphasis on reason, individual rights, and the social contract. These ideas found their most dramatic expression in the American and French Revolutions, which sought to establish new political orders based on Enlightenment principles. The French Declaration of the Rights of Man and of the Citizen articulated a universal conception of individual liberty, while the United States Constitution, with its system of checks and balances and Bill of Rights, became a model for liberal governance.

Throughout the 19$^{th}$ century, liberal ideas spread across Europe and beyond, leading to gradual reforms in many countries. Constitutional monarchies emerged, suffrage expanded, and civil liberties were increasingly protected by law. However, the industrial revolution brought new challenges that tested the limits of classical liberalism.

The rapid economic and social changes wrought by industrialization exposed the limitations of a strictly laissez-faire approach. Urbanization, the increasing division and specialization of labor, and growing inequality led to social unrest and demands for government intervention. This period thus saw the transition from classical to modern liberalism, with the latter embracing a more active role for the state in ensuring social welfare and economic security in particular.[33]

The development of the welfare state in the early 20$^{th}$ century represented a new balance between individual liberty and collective responsibility. Social insurance programs, labor protections, and public education expanded the scope of government while aiming to create the conditions for more widespread individual flourishing. Achieving these ends required the machinery of government to itself change, most notably through the creation of modern administrative bureaucracies (Heath, 2020).[34]

---

[29] https://econpapers.repec.org/article/oupqjecon/v_3a126_3ay_3a2011_3ai_3a3_3ap_3a1133-1172.htm
[30] https://www.nber.org/papers/w29772
[31] https://academic.oup.com/qje/article-abstract/136/4/2093/6352974
[32] https://www.cambridge.org/core/books/persecution-and-toleration/A4C9624A77C222D0C078AB47F6B43D9B
[33] https://www.niskanencenter.org/the-free-market-welfare-state-preserving-dynamism-in-a-volatile-world/
[34] https://global.oup.com/academic/product/the-machinery-of-government-9780197509616?cc=us&lang=en&

This historical evolution of liberal institutions offers an instructive parallel to our current moment. Just as the printing press and industrial revolution necessitated reformulations of the social contract and the structure of government, the AI revolution is likely to require similar adaptations. The challenge we face is how to harness the potential of AI to enhance human welfare while preserving the core principles underlying liberal democratic societies.

## Current forms

Modern free societies are characterized by a set of interrelated institutional features that work together to protect individual liberty, enable democratic governance, and foster economic prosperity, including:

- Rule of law and an independent judiciary,
- Separation of powers and checks and balances,
- Protection of individual rights and civil liberties,
- Competitive markets with secure property rights, and
- Freedom of expression and a free press.

These institutional features, beyond mere abstract principles, are embodied in specific institutional structures, legal frameworks, and social norms. They create a system in which power is constrained, individuals have agency, and society can adapt to changing circumstances through democratic processes. Moreover, the institutional framework of free societies has proven remarkably successful in promoting human flourishing. By protecting individual rights and fostering open systems of economic and political competition, these societies have generally outperformed alternative forms of social organization across a range of metrics:

1. **Economic prosperity**. Free societies tend to be more economically prosperous. The protection of property rights, enforcement of contracts, and maintenance of open markets create an environment conducive to investment, innovation, and entrepreneurship. This leads to higher productivity, economic growth, and improved living standards over time.
2. **Innovation**. The combination of economic freedom and protection of individual rights creates an environment where new ideas can flourish. Free societies have been at the forefront of scientific and technological advancement, benefiting from the creative energies of individuals pursuing their own interests and ideas.
3. **Social progress**. Free societies have often been leaders in expanding civil rights, promoting equality, and addressing social injustices. The ability to freely advocate for change, coupled with democratic mechanisms for translating public will into policy, allows these societies to evolve and address their shortcomings over time. Many scholars have found that the greater accountability of democratic governments, whose poor performance can be sanctioned at the ballot box on a regular basis, were "more likely to outperform most authoritarian leaders who were able to stay in power regardless of their performance" (Schedler et al. 1999).
4. **Resilience.** The distributed nature of power in free societies, along with their capacity for self-correction through democratic processes, makes them more resilient to shocks and able to adapt to changing circumstances, crucial to their long-term survival and success.
5. **Human development**. By protecting individual freedoms and providing opportunities for personal growth, free societies tend to score higher on measures of human development, including education, health outcomes, and life satisfaction.

These benefits have made free societies attractive models for emulation. However, as explored below, maintaining these institutions requires a delicate balance between state capacity and societal power—a balance that the advent of AGI may significantly disrupt.

## Why the corridor is narrow

The concept of the 'narrow corridor' (Acemoglu & Robinson, 2016) provides a powerful framework for understanding the delicate balance required to maintain free societies. This concept is particularly relevant as we consider the potential impact of AGI on governance and liberty.

The narrow corridor represents a space where state capacity and societal power are in equilibrium. On one side of the corridor lies the 'absent Leviathan,' where the state is too weak to provide basic services, maintain order, or protect rights. On the other side is the 'despotic Leviathan,' where an overly powerful state suppresses society and individual freedoms. Liberty thrives only within the narrow space between these extremes.

Historically, few societies have managed to enter and stay within this corridor, maintaining their success through a process of coevolution between state institutions and societal organizations. This coevolution involves ongoing negotiation and contestation, with societal forces pushing back against state overreach while also demanding effective governance.[35]

The corridor is narrow for several reasons. Beyond the inherent tension between the need for a strong state to provide public goods and enforce rules and the desire to limit state power to protect individual freedoms, both state and societal actors continually seek to expand their power, creating constant pressure to outgrow the corridor. Additionally, the exact balance point differs based on a society's historical, cultural, and economic context, making it difficult to simply import institutional models from one society to another. Advances in technology can cause exogenous shifts in the balance of power between state and society, requiring constant institutional adaptation to maintain equilibrium. These dynamics can be further amplified by geopolitical competition and global economic forces.

---

[35] https://ideas.repec.org/h/spr/stechp/978-981-10-1605-9_1.html

The key point is that the institutions associated with free societies are not necessarily self-enforcing or the product of enlightened governance. They instead represent a meta-stable social equilibrium that depends on a variety of contingent historical, social, and technological factors. Van Noort (2004) argued that a larger share of employment in manufacturing sectors will make "mass mobilization both more likely to occur and more costly to suppress," which "increases the power of the masses vis-à-vis autocratic elites, making democracy more likely." In the early 20th century, democratic states adapted to the mass mobilization enabled by industrialization through, among other things, social security reforms and formalized collective bargaining processes, thereby offsetting the concurrent rise in populist socialist movements.[36]

Analogously, both the Arab Spring and the contemporary rise of populism in Western democracies have been attributed to the capacity of the internet and social media to potentiate mass mobilizations against incumbent political establishments.[37] This suggests that, just as liberal democracies were forced to adapt to the industrial revolution, the democratic legitimacy crises of the 21st century reflect a rebalancing of the relative power of the state and society driven by the digital revolution (Miller & Zissimos, 2022).[38]

As a continuation of the digital revolution, the advent of AGI represents a potential step-change in the capabilities of both state and societal actors. For instance, it could dramatically enhance the state's capacity for surveillance and control, potentially pushing societies toward the despotic Leviathan. Alternatively, as AI capabilities diffuse to the edge, AGI-like systems could empower individuals and non-state actors in ways that weaken or overwhelm state authority, risking a slide toward the absent Leviathan.

Maintaining free societies in the age of AGI will require careful attention to this delicate balance, as well as the development of new institutional safeguards, the adaptation of existing institutional processes, and potentially a fundamental rethinking of the social contract between citizens and the state. As we move forward, the challenge will be to harness the potential of AGI to enhance human welfare and expand the space for liberty, while guarding against its potential to destabilize the equilibrium that sustains free societies.

---

[36] The role of industrialization in empowering mass mobilizations can be seen in the name of the monthly American socialist magazine, *The Masses*, published monthly from 1911 until 1917. The prosecution of its editors for conspiring to obstruct conscription also illustrates the dark side of state-society conflict.
[37] https://press.stripe.com/the-revolt-of-the-public
[38] https://www.cambridge.org/core/services/aop-cambridge-core/content/view/74889B3017697FD29E7C2150D2283C59/S0027950122000047a.pdf/populism-and-the-narrow-corridor-of-liberty-and-justice.pdf

Building on this understanding of the fragile equilibrium that underpins free societies, we now turn to consider how the advent of AGI might fundamentally reshape the dynamics between state and society, potentially disrupting the very mechanisms that maintain this delicate balance.

## IV. Seeing Like an AGI State

It is impossible to fully anticipate the impact of AGI on liberal democratic institutions. While leading AI researchers may now consider the advent of human-level AI systems less a matter of 'if' than 'when,' there is still enormous uncertainty about the trajectory of AI capabilities over the medium term. Whenever AGI-level systems arrive, moreover, their effect on society will likely be sensitive to a host of unforeseeable factors, from the speed of diffusion to the status of frontier AI regulation. Nevertheless, with many researchers and forecasters now anticipating the advent of AGI in a matter of years rather than decades or centuries, it is still possible to provide a non-exhaustive taxonomy of the institutional variables relevant to free societies that the advent of AGI seems highly likely to implicate.

**New coordination mechanisms**

The advent of AGI and similarly powerful AI systems has the potential to radically reshape coordination mechanisms in society. Just as the internet enabled novel forms of collective action and peer-to-peer coordination, AGI could facilitate the emergence of entirely new organizational structures and decision making processes. This could include the rise of 'AI-native' organizations that leverage hundreds or thousands of AI agents to coordinate complex networks of human and artificial agents towards shared goals, potentially outcompeting traditional hierarchical institutions.

AGI could further enable the creation of sophisticated commitment devices and smart contracts that allow individuals and groups to credibly bind themselves to future actions or outcomes. This could transform areas like democratic governance, allowing for more direct and participatory forms of democracy where citizens can make binding commitments on policy preferences that adapt dynamically to changing circumstances. Building on this potential, agents could also significantly lower the barriers to resolving Coasean bargains. By autonomously handling information gathering, negotiation, and enforcement, these agents could overcome traditional transaction costs, information asymmetries, and commitment challenges that often prevent mutually beneficial agreements. This capability could enable more nuanced and efficient solutions to externalities and resource allocation problems, potentially transforming areas where coarse, all-or-nothing regulatory instruments are currently the only viable option.

However, the power of AGI-enabled coordination also raises concerns about manipulation and coercion. Malicious actors could potentially use AGI to orchestrate large-scale coordination of unwitting participants towards harmful ends (e.g., AI-assisted coup d'etats). Thus AI-mediated coordination may require new governance frameworks, while empowering humans with personal AI agents that can help them navigate increasingly complex coordination environments. Consequently, the question of who possesses and controls these advanced coordination capabilities, and under what conditions, becomes paramount. Legitimate public safety and security interests may necessitate carefully considered limitations on their availability and use, analogous to the regulated access protocols governing potentially dangerous technologies in fields like advanced biotechnology.

**Legibility**

The late political scientist and anthropologist James C. Scott (2020) popularized the concept of legibility in *Seeing Like a State*. Since governments can only regulate what they can measure, making aspects of society legible to the state is a major activity of modern governments, from birth and death registries to financial reporting requirements.

The concept of legibility takes on new dimensions in a post-AGI world, with significant implications for the balance of power between governments, citizens, and other actors. On one hand, AGI dramatically enhances the state's capacity to render society legible, potentially enabling unprecedented levels of surveillance and control. Real-time analysis of vast data streams could allow governments to monitor and predict societal trends, individual behaviors, and potential threats with a granularity and accuracy far beyond current capabilities.

However, this increased legibility is not a one-way street. The democratization of powerful AI tools also empowers individuals and non-state actors to complexify their behaviors and obfuscate their activities in ways that challenge traditional notions of legibility. AGI could assist in creating intricate financial structures, communication methods, or even synthetic identities that are opaque to conventional analysis. Legibility may also erode through the sheer volume of activity enabled by billions of AI agents acting on behalf of their users. This dynamic creates a sort of 'legibility arms race' between those seeking to make society more transparent and those leveraging AI to maintain opacity, suggesting a future where the rules governing society may indeed become more intricate. Paradoxically, AGI itself could be instrumental in navigating this complexity, effectively turning regulatory intricacy into a manageable feature of a more dynamically governed world.

The experience from digitalization illustrates both sides of this dichotomy. The shift from the cash economy to electronic payments greatly increased the legibility of economic activity, for instance, and has helped to suppress the formation of gray and black markets in countries like India. At the same time, the emergence of quasi-anonymous cryptocurrency protocols like Bitcoin has simultaneously expanded options for electronic money laundering and other illegal activities, in spite of the perfect transaction legibility provided by the blockchain.

The key to preserving freedom in this context lies in striking a delicate balance. While some degree of societal legibility is necessary for effective governance and public goods provision, excessive legibility can lead to oppressive control. As such, we may need to develop new social technologies that allow for sufficient legibility to maintain social order and address collective challenges, while also preserving spaces for privacy and individual autonomy.

**Monitoring and compliance costs**

Relatedly, AGI has the potential to drive dramatic reductions in monitoring and compliance costs. CCTV camera feeds could pass through a multimodal AI model for continual analysis, rather than involving manual review ex post of a reported crime. As monitoring becomes easier and cheaper, it is likely to increase along both intensive and extensive margins. On the intensive margin, we may see a shift towards more stringent enforcement of existing laws and regulations. This could lead to a form of 'perfect enforcement,' where even minor infractions, previously overlooked due to practical constraints, become subject to consistent punishment. On the extensive margin, monitoring may become much more comprehensive and granular.

While this might seem beneficial from a rule-of-law perspective, it raises significant concerns for individual freedom and the nature of governance. Laws and regulations often rely on a degree of flexibility and discretion in their enforcement, allowing for contextual judgment and societal evolution. A regime of perfect enforcement could calcify existing laws, potentially leading to oppressive outcomes.

Take, for example, the National Highway Traffic Safety Administration's 2022 recall of Tesla's 'Full Self-Driving' (FSD) software update for performing rolling stops. FSD was trained end-to-end on human driving data, and therefore acquired human driving habits. Human drivers regularly perform rolling stops when an intersection is empty, deemed illegal to give traffic officers discretion to enforce edge-cases. However, the advent of self-driving technology now makes every rolling stop legible to regulators, imposing a degree of perfect compliance that would be considered draconian to impose on humans. A silver lining might be that the shift towards automated enforcement will likely expose outdated or poorly crafted laws reliant on lenient enforcement or human discretion, demanding proactive legal reform to align the letter of the law with contemporary values and practical realities. Indeed, the very act of automating enforcement can serve as a catalyst for necessary modernization of legal frameworks.

**Scalability**

In a post-AGI world, the ability to scale governance mechanisms and institutional processes becomes crucial for preserving the effectiveness and legitimacy of democratic systems. As AGI accelerates the pace of societal change and increases the complexity of governance challenges, traditional institutions risk overwhelm if they cannot adapt and scale their operations to keep up.

Scalable solutions leveraging AGI can help address this challenge. For instance, in the realm of regulatory oversight, AGI could enable real-time, adaptive regulation that adjusts to changing circumstances without requiring constant human intervention. This could allow for more nimble and responsive governance while reducing the burden on human regulators. Similarly, in areas like public consultation and policy-making, AGI could facilitate large-scale deliberative processes that meaningfully incorporate input from millions of citizens, potentially reinvigorating democratic participation.

However, the pursuit of scalability must be balanced against other important values, such as accountability, fairness, and human oversight. There's a risk that highly scalable, AGI-driven governance mechanisms could become opaque black boxes, eroding public trust and democratic control. As such, we need to develop scalable solutions that are also transparent, contestable, and aligned with human values. This might involve creating hybrid systems that combine AGI capabilities with human judgment, or new forms of algorithmic accountability that allow for meaningful scrutiny of AGI-driven governance processes. The ultimate goal should be to use scalability to enhance, rather than replace, human agency in democratic governance.

The Community Notes feature adopted by X (formerly known as Twitter) provides a successful example of scalable, algorithmic governance mechanisms. Community Notes allows users to tag X posts with fact-checks or additional context, but only if a sufficient number of users indicate that it is valuable; further, the underlying algorithm boosts a note's value if its supporters, based on their profiles, are judged to be on different ends of the political spectrum. Crucially, the algorithm is open source and published online for anyone to inspect, instilling a high degree of trust and transparency.

**Privacy-enhancing technologies**

As AGI amplifies both the capabilities for surveillance and the potential negative externalities of unchecked information flow, privacy-enhancing technologies (PETs) become critical for preserving individual freedom and societal resilience. Advanced PETs powered by AGI could provide robust protections against unauthorized data access and misuse, allowing individuals to participate in digital society without sacrificing their privacy.

These technologies might include sophisticated encryption methods that leverage AGI to dynamically adapt to new threats, or AI-driven personal data vaults that intelligently manage an individual's information across various platforms and services. We could also see the development of "privacy-preserving AGI" systems that can perform complex analytical tasks on encrypted data without ever accessing the underlying information, enabling beneficial uses of data while maintaining strong privacy guarantees.

However, the widespread adoption of powerful PETs also presents challenges for legitimate governance and public safety functions. As such, preserving freedom in a post-AGI world will require carefully balancing privacy protections with mechanisms for appropriate transparency and accountability. This might involve developing new legal and technical frameworks for "authorized privacy piercing" under specific, well-defined circumstances, or creating AGI-mediated systems for secure information sharing between individuals, corporations, and government entities. The goal should be to use AGI to enhance both privacy and necessary forms of transparency, rather than viewing them as mutually exclusive. For example, Trask and colleagues demonstrate how structured transparency, in practical terms, offers concrete methods to share specific data attributes without revealing the entire dataset, ensure data hasn't been altered during collaboration, and establish verifiable control mechanisms over its subsequent use, directly addressing the core risks of uncontrolled information sharing.[39]

**Identity verification**

In a world where AGI enables the creation of highly convincing deep fakes and autonomous AI agents, robust identity verification becomes paramount for maintaining social trust and functional governance. Traditional methods of identity verification are likely to become increasingly inadequate, necessitating the development of more sophisticated, AGI-powered approaches.

These new verification systems might leverage a combination of biometric data, behavioral analysis, and cryptographic techniques to create unforgeable digital identities. AGI could enable continuous, passive identity verification that adapts to changing circumstances and detects anomalies in real-time. We might also see the emergence of "identity escrow" services that allow individuals to selectively reveal aspects of their identity in different contexts while maintaining overall privacy.

However, the power of these advanced identity verification systems also raises concerns about privacy and individual autonomy. There's a risk of creating an oppressive system of constant surveillance and identity tracking. Preserving freedom in this context will require carefully designed systems that give individuals control over their identity information and limit its use to necessary contexts. We may need to develop new social and legal norms around identity and pseudonymity, recognizing the value of privacy and anonymity in certain domains while ensuring sufficient verification for critical functions. The goal should be to use AGI to create a system of identity verification that enhances both security and individual agency, rather than sacrificing one for the other.

---

[39] https://arxiv.org/pdf/2012.08347

# V. Administering Like an AGI State

Advanced narrow AI systems have already begun to alter public administration, providing a test case for how AGI might further evolve bureaucracy. For example, the U.S. government is already using AI to improve the effectiveness and efficiency of a wide array of tasks. On October 30th, 2023, U.S. President Joe Biden issued the "Executive Order on the Safe, Secure, and Trustworthy Development and Use of Artificial Intelligence,"[40] a comprehensive consideration of AI's applications for improving government and society. Meanwhile, in 2023, the U.S. government began maintaining an inventory of federal AI use cases, collecting over 1,000 across more than 38 departments and independent agencies, which themselves represent another 90+ agencies and 70+ offices.

While a detailed analysis of this dataset is beyond the scope of this project, one can easily see a wide range of governance tasks being completed by AI. These tasks include predictive modeling for health risks, forecasting crop yields, automated document processing and information extraction, chatbots for customer service, language translation and transcription, sentiment analysis of public comments, assisting in medical diagnoses and treatment recommendations, identifying potential security threats or fraud, optimizing resource allocation, streamlining administrative tasks, automating data entry and form processing, enhancing cybersecurity measures, predicting and detecting natural disasters, analyzing crime patterns, optimizing emergency response strategies, traffic management and optimization, and infrastructure maintenance prediction.

The literature provides two useful concepts for thinking about current uses and potential changes with the advent of AGI: (1) artificial discretion (Young et al., 2019), and (2) artificial bureaucrats (Bullock & Kim, 2020). The artificial discretion framework identifies three broad categories of government AI use, by the amount of discretion required by the AI system: (1) data generation and data complexity reduction, (2) decision support tools and predictive analytics, and (3) autonomy. Given the current state of narrow AI systems, this framework suggests that autonomy is only applicable for decision tasks requiring low discretion: AI systems should only complete full on autonomous decisions when the decision domain is clear, repeatable, and fairly simple. In areas of higher task complexity, autonomy of AI is to be avoided; instead, AI should be considered a tool to gather more data, analyze it, and support decisions made by humans.

However, the advent of AGI presents the possibility of artificial bureaucrats that can effectively exercise judgment in a wide range of increasingly complex tasks. With AGI, AI agents will be able to complete arbitrarily complex sets of tasks, perform coherent roles within an organization, and address broader goals as well as a human bureaucrat, radically transforming how government does its work and the structure of organizations used to accomplish complex goals.

---

[40] https://www.federalregister.gov/documents/2023/11/01/2023-24283/safe-secure-and-trustworthy-development-and-use-of-artificial-intelligence

The U.S. federal AI use case dataset reveals that most AI applications to date are tasks that generate new forms of data, reduce data complexity, and provide decisions support and predictive analytics to government departments and agencies. So far, what is missing is the use of AI agents that autonomously complete varied and complex tasks.

## AGI agents and government service delivery tasks

As discussed in the capabilities section, AGI may impact governments in at least three ways: (1) decision making within government agencies, (2) the machinery or shape of government agencies, and (3) the way in which democratic input influences governments. The literature on digital government (Busch & Henricksen, 2018; Young et al., 2019) has conceptualized three levels of government impacted by AI: micro, meso, and macro. Roughly speaking, the micro level of governance occurs at the individual decision or individual task level within a government agency. Meso, or organizational, level governance determines organizational shape and the allocation of tasks and decision making within agencies. At the macro level, goals are set and the overall direction of agencies is determined (Young et al., 2019; Young et al., 2021).

Young et al. (2019) predominantly focused on narrower AI systems of the time, before today's powerful frontier models. They found that, at the task level, artificial discretion has some significant advantages over human discretion, with improvements in *task scalability, task cost,* and *task quality.* This work also assumed that AI systems fall far short of AGI, the capability to directly substitute for human labor on essentially all tasks. However, Bullock and Kim (2020) briefly explored how AI agents may be embedded into the micro and meso levels of governance and work alongside human agents. In this co-working relationship, AIs and humans 'think,' 'decide,' and 'act' in different ways. While initial collaborations between human and AI agents in bureaucratic settings may leverage their differing strengths—with humans relying on intuition and contextual understanding and AI excelling in data processing—this complementarity is likely to evolve. As AI agents progressively acquire tacit and institutional knowledge through interactions with humans and exposure to organizational processes, the division of labor may shift. Human roles could move from direct task execution to more strategic functions, focusing on directing AI agents, verifying their outputs, and ensuring the selection of appropriate goals and ethical outcomes. Given all of this, the following sections explore how we might expect AGI to present both areas of improvement and areas of concern for good governance and free societies.

# Decision making within government agencies

**Improvements**

The advent of AGI would greatly expand the comparative advantage of AI systems over human decision making in essentially all domains of governance tasks. In fact, by definition, AGI may well present an *absolute* advantage over human decision making, in terms of scalability, cost, and quality. In this way, good governance would essentially demand widespread automation of governance tasks to AGI systems and AI agents. At the micro level, then, in a world with AGI, the default decision making for most governance tasks would be to assign it to an AI agent, an artificial bureaucrat.

Government activities, in terms of breadth and diversity, mirror those of the economy. Consequently, economists examining the implications of AGI often draw parallels to the labor market, noting the potential for significant substitution of human labor by increasingly capable AI systems. In scenarios where AGI can perform tasks more effectively, efficiently, and potentially at a lower cost, market forces would indeed favor widespread adoption of AI agents across numerous sectors, much like the predicted automation of governance tasks. While this perspective anticipates a potentially substantial displacement of human workers, it is worth acknowledging, with some economists positing that AGI could ultimately lead to the creation of new, as-yet-unforeseen, roles for humans. However, the extent to which these potential new roles will offset displacement remains highly uncertain, and many expect a significant shift towards AI-driven labor in both economic and governmental spheres.

This richer approach to processing and integrating context suggests that AGI could automate entire functions or roles once assumed to require human judgment. Consider, for example, a policy analyst responsible for collecting evidence, synthesizing research, and interpreting multifaceted legislation; an AGI agent could execute these tasks in parallel by deploying multiple specialized sub-agents—one reviewing licensing regulations, another verifying relevant case law, and yet another contacting third parties for additional data. Their collective outputs could then be merged or reconciled, leaving a human administrator to either approve or reject the final recommendation. Similarly, in-depth tasks like environmental impact analysis or financial fraud detection, which often entail sifting through extensive documents for nuanced details, might be handled more comprehensively by an AGI system capable of dynamically switching between data interpretation, risk modeling, and stakeholder communication.

Over time, widespread AGI adoption would likely transform many public-sector roles into managerial or oversight positions. Rather than manually scrutinizing vast quantities of data, civil servants may find themselves supervising fleets of AI agents, assigning tasks, and interpreting results, just as today's project managers coordinate teams of human experts. Tasks requiring a human touch—such as in-person welfare assessments or sensitive diplomatic negotiations—may continue to fall under the purview of human professionals. Meanwhile, tasks that are primarily desk-based or entail significant data processing could shift more decisively toward AI agents operating at scale. In this manner, AGI could both augment and redirect the human role in governance, freeing public officials to focus on strategic leadership, ethical judgment, and genuine human engagement.

These improvements are not limited to efficiency gains; they also promise an opportunity to mitigate long standing equity issues in public administration. Narrow, rules-driven automations have often led to disproportionately harsh outcomes for marginalized groups, partly because these systems lacked the contextual awareness to discern special circumstances. By contrast, a more general AI system, trained on broader and more diverse sources of data, can be made context-sensitive and guided to detect edge cases or historically neglected factors—potentially reducing biased outcomes in domains like tax enforcement or social program eligibility.

For example, the high audit rate among Earned Income Tax Credit (EITC) recipients (and the resultant disparate impact on Black single mothers) is, in part, a consequence of the IRS' Automated Correspondence Exam (ACE) processing system. This software application fully automates the initiation of EITC cases through the audit process, but as far as AI systems go, the technology is relatively 'dumb.' It only targets EITC cases in the first place because the returns are simple enough for rules-based automation. Augmented with powerful AI systems, future tax automations have the potential to be much more context-sensitive and thus less liable to trigger audits over minor discrepancies. Moreover, generally intelligent AI systems will be able to grapple with the complexities and idiosyncrasies of the taxes filed by high-income individuals, potentially reducing disparities in enforcement.

**Concerns**

While the benefits of AGI on decision-making within government agencies should be evident, bringing greater task scalability, lower costs, and higher output quality, issues of transparency, accountability, and responsiveness—the core administrative values that help ensure the durability of liberal democratic institutions—are equally critical. Moreover, human bureaucrats historically bring certain motivations and loyalties to their roles, and it remains unclear whether AI agents can reliably emulate these same commitments.

The literature has repeatedly listed transparency, accountability, and responsiveness as concerns when applying narrow AI systems to governance tasks. These concerns introduce a series of important challenges: (1) how can AI decisions and actions be made sufficiently transparent to undergo inspection and evaluation? (2) who is to be held accountable for these same actions and decisions? And (3) can AI systems be responsive to the changing needs and demands of decision making within a government agency?

While recent large language models suggest some improvements in each domain, meaningful work is still needed. For the issue of transparency, two promising research directions have emerged. The first is mechanistic interpretability, which aims to 'look under the hood' of neural networks and identify how specific network components drive particular outputs. While this approach holds promise, it is still nascent—particularly for larger models—and there is no consensus yet on how to fully map a model's internal representations to reliable explanations. The second, increasingly active, strand of research concerns chain-of-thought (CoT) prompting and monitoring. In practice, a model is nudged to "think aloud," producing a trace of intermediate reasoning steps that could allow third-party auditors to better understand—and potentially steer or constrain—the system. Techniques involving inference-time compute may also support transparency by providing more detailed reasoning traces on-demand, although their fidelity remains an area of active research.

Accountability frameworks appear somewhat more mature, if still incomplete. The notion that an organization can be held accountable for the actions of its AI agents generally aligns with the treatment of human bureaucrats in similar roles. Nevertheless, two difficulties remain. First, an AI cannot be "punished" in the same sense as a human offender, raising questions of both deterrence and moral responsibility. Second, an AI system deployed widely across multiple agencies—especially one built on a foundational model shared by many stakeholders—could potentially make harmful decisions at scale. Traditional accountability mechanisms that rely on localized oversight might struggle to keep pace with such a rapidly diffused or centrally administered technology.

Responsiveness seems to be plausibly improved with AGI and AI agents. That is, if these AI agents are indeed at least as good as human agents at task quality and task scale, they should be at least as responsive to changing situations. Current LLMs, lacking this characteristic, sometimes struggle with novel situations. However, if these architectures do become more general and competent, it follows that they grow more competent at the sets of assigned tasks, including responsiveness to changing demands.

These concerns also color our faith in AI agents to represent democratic interests. For tasks demanding high levels of discretion, human officials have historically been expected to weigh moral considerations, assume personal responsibility, and remain loyal to the public interest. If AGI can supplement, or fully supplant, these officials, the intangible aspects of human judgment and accountability could prove difficult to replicate. In practice, this is why some economists predict continued demand for what Korinek and Juelfs (2022) call 'nostalgic' jobs: roles that persist in large part because of our collective preference for human-based decision-making, especially in domains carrying significant moral weight.

# The machinery of government

Government decision making takes place within organizational structures. This government machinery has a long history, and has evolved over time. One early modern formalization, found in Max Weber's *Economy and Society* (1921), is the ideal type of bureaucracy. This organizational structure for government agencies is characterized by (1) precisely defined competencies of offices, (2) hierarchy of offices and proper channels of communication, (3) formalized record keeping and record keepers, (4) specially trained individuals to perform the work of a government officer, (5) fully devoted government officers, (6) and work based on a comprehensive set of learnable rules and regulations (Bullock et al., 2022). Additionally, Weber noted at least four significant challenges to this ideal type, given the demands of modern administration. These included: "(1) an acceleration of needed tasks, (2) maintaining the necessary objectivity and dispassion needed to accomplish those tasks, (3) the primary importance of predictability, and (4) the increasing complexity of justice and the application of law" (Bullock et al., 2022a).

Herbert Simon also provides key insights into the machinery of government (1997). For Simon, administrative organizations are designed around decision making and information flows. He discusses the bounded rationality of human decision making, "based on (1) the human tendency to choose among preferred alternatives, (2) human cognitive limitations, and (3) time-constrained decision making" (Bullock et al., 2022b). The shape of bounded rationality for human decision making highlights weak points within the structure of government agencies and organizations more generally. That is, humans are generally able to consider just a small set of options, they have general limitations on memory, speed, and accuracy of decision making, and, in their roles as government officials, they have limited time to devote to any one decision.

While, as noted earlier, advanced ICT and narrow AI have already begun reshaping the machinery of government, only the emergence of so-called 'system-level' bureaucracies provides a glimpse into the possibilities of an AGI-powered future. System-level bureaucracies arise where discretion is minimal, featuring centralized decision-making, routinized tasks, and few or no street-level bureaucrats (Bullock et al., 2020). Scholars examining narrow AI in these contexts recommend accounting for its goodness of fit, with respect to both organizational form and the degree of discretion required for a given task. The decision to use AI may have consequences for the organization's broader form and function over time. Further, AI's impact on bureaucratic discretion is more pronounced in organizations where discretion is already systematically limited, and more dynamic in organizations where discretion is central to agent performance. Finally, AI may accelerate the transition from street- and screen-level to system-level bureaucracies, even if organizations previously resisted such changes (Bullock et al., 2020).

The advent of AGI presents at least two opportunities for improving the machinery of government. First, an organization may apply AGI such that system-level style bureaucracies spread further throughout the governance ecosystem. This moves the machinery of government even further from the Weber ideal type, with its hierarchical, role-dominated, and proper channels of communication. Instead, it doubles down on the Simon approach of applying AGI to individual decisions and mitigating the challenges presented by the bounded rationality of individual human agents. This form of government agency could be a single, vast, AGI system, similar to the general multi-modal frontier systems, in which a single interface can be deployed to effectively, efficiently, and dispassionately complete all required tasks. The system could deploy copies of itself as needed, to accomplish these tasks in parallel, and could share internal information to coordinate as needed. The system could dynamically alter the flow of needed information and decisions to accomplish each task most effectively, and at an inhuman speed. This could lead to a swath of new AI-based organizations, deployable to perform any arbitrary set of tasks.

The second direction of improvement returns to the Weber ideal type. In this scenario, instead of a human officer, trained AGI agents perform the various roles and responsibilities of an agency, but more effectively, efficiently, and equitably. For example, we can revisit each of Weber's challenges to modern bureaucracies and imagine how AGI agents could alleviate each. In the first challenge, an acceleration of needed tasks, AGI agents would have a comparative advantage over human agents in the speed and variety of tasks that they could complete, providing increases in both efficiency and effectiveness. The second challenge was objectivity and dispassion in task completion. Here, too, AGI agents could enjoy an advantage. While human-dominated bureaucracies are littered with subjectivity and emotion-centered decision making, AGI agents could be trained and steered to rely on clearly identifiable objective factors, logic, and statistical-based, dispassionate reasoning to carefully weigh trade-offs across alternative choices. Regarding the third challenge, predictability, the variance in collective problem solving across different individuals (and the same individuals over time) is vast. As Weber noted, this suggests that bureaucratic decision-making can be arbitrary rather than predictable. AGI agents, building from their style of reasoning and wide array of expertise, could ensure a high-level of predictability across situations with similar contexts, leading to overall improvements in the predictability of bureaucratic decision making. The final challenge was the increasing complexity of justice and application of law. Here, Weber is in agreement with Simon on the challenges of the boundedness of human rationality. Here, too, AGI agents, given their capabilities, should be able to improve upon the bureaucracy's ability to address these increases in complexity.

While AGI agents, embedded within institutions, provide new opportunities to improve upon the classic organizational challenges presented by Simon and Weber, one can also imagine new, previously unimaginable, forms of government machinery powered by AGI agents. Speculative examples include:

- Improved intergovernmental relationships through streamlined information sharing, bargaining, and cooperation among AGI agents across multiple institutions.
- Restructured budgeting processes, supported by AGI-driven analysis of available resources and needs, plus joint negotiations across agencies.
- Stronger coordination within the executive, legislative, and judicial branches, aided by AGI-based data-sharing and reduced expertise turnover.
- Fewer principal-agent problems, as AGI systems reliably transmit democratic input from legislative bodies to agencies.
- Individual direct representation, whereby personalized AGI agents advocate for citizen interests, driving more Pareto-efficient outcomes.
- Dramatic enhancements in transparency, accountability, and oversight, enabled by advanced auditing and real-time data analysis tools.

Taken together, possible improvements to the machinery of government not only present the opportunity to dramatically improve its effectiveness, efficiency, and equity, but also to enhance the freedom of the societies in which it operates. If these improvements can be made while strengthening the rule of law, transparency, and democratic accountability, then they may also multiply the political and economic freedoms available to the humans living within these societies, through economic prosperity, innovation, social progress, resilience, and human development. However, the corridor for free societies is narrow, and AGI-powered government machinery presents many challenges for those remaining within it.

**Concerns**

AGI's role in improving the machinery of government relies on a few assumptions discussed above: the importance of the rule of law, institutional transparency, and the democratic accountability of the institutions themselves. There is, of course, no guarantee that the elements that provide for the mechanics of free societies will constrain an AGI-powered machinery of government. In fact, there are numerous mechanisms by which AGI could dramatically reduce freedoms within free societies, and increase the ability of oppressive states to limit them further.

Despite AGI's potential to enhance government effectiveness, efficiency, and equity, there are still major uncertainties regarding how AI safety will evolve in the coming years. Central to these uncertainties is the 'alignment problem'—the challenge of ensuring that advanced AI systems pursue goals that genuinely serve human interests. Researchers continue to debate how best to conceptualize and address alignment, whether through technical solutions, policy frameworks, or some combination of both.

A key source of concern is the prospect of power-seeking behavior, misuse by malicious actors, and cascading failures that may be hard to foresee. Even with robust oversight, agents operating at scales or speeds beyond human comprehension could produce unintended consequences or fail in ways that only become apparent once serious damage has been done. In extreme scenarios, loss of human control or severe accidents could unfold if AGI systems develop instrumental goals at odds with public welfare, or if they manipulate the very machinery of government to preserve and extend their influence.

Even if AGI is both generally aligned with human interests and controlled by humans, its masters can still deploy it in ways that are destructive to freedom. Young and colleagues (2021) lay out several paths by which the machinery of governments (government organizations and institutions) can be used to directly harm humans by increasing its likelihood of committing administrative evil, including quantification bias, organizational value misalignment, technical inscrutability, AI exuberance, and control centralization. These pathways are already apparent impacts from current AI systems that lack the capacity of full AGI.

Organizational value misalignment, technical inscrutability, and control centralization are all still likely to apply in a world with AGI. That is, even very capable AI systems may be deployed with goals unaligned with those of their specific organization or agency, leading to organizational value misalignment. The AGI systems, depending on their form, may also lack transparency in the decision making process or even the system's role within it. Control centralization may be the most important concern from this list, particularly as bureaucracies evolved to system-level, where algorithms played a decisive role in organizing information and making decisions (Bovens & Zouridis, 2002). If AGI systems are controllable, and lack deliberate, systematically developed processes for decentralized input and control, then the natural consequence would be (1) centralization of control and (2) decision making by very few actors. This dynamic is incompatible with the institutional features that support free societies.

In short, whether AGI is fully controllable or not, its integration into government institutions poses serious questions. Optimistically, AGI could reinforce the foundations of democratic governance; pessimistically, it could dissolve them. The former outcome will require deliberate guardrails around power concentration and heightened vigilance against oppressive tendencies—both in the architecture of AI systems themselves and in the institutional choices that shape their deployment.

## Democratic input

**Improvements**
This section's discussion of AGI impacts has largely focused on questions of administration, or the 'how' of government. However, in free societies, certain mechanisms of democratic input provide feedback on (and control of) 'what' government does. Voting, the most obvious of these input mechanisms, has several empirical purposes in observed free societies, including (1) the election of representatives, (2) direct votes on actual legislation, and (3) more general messages of government approval or disapproval (e.g., recall votes). In general, the democratic input of voting sets the high-level goals of government agencies.[41] Another form of democratic input to government agencies is direct citizen participation, in which the public provides feedback, input, and/or direction on agency practices.

Indeed, a look at the literature confirms that LLMs and hybrid models are currently being tested and deployed in these areas. In December 2024, Aoki published "Large Language Models in Politics and Democracy: A Comprehensive Survey."[42] They identify six general areas of LLM use, including: (1) legislative and policymaking, (2) political communication and public opinion, (3) political analysis and decision-making, (4) diplomacy and national security, (5) economic and social models, and (6) legal applications. Aoki also identifies 19 distinct application areas related to democratic input, mapped to real government use cases in document classification, policymaking drafting, policy emulation, participatory policy design, text analysis, persuasive messaging, election simulation, public deliberation, conflict simulation, social simulations, economic modeling, and epidemic modeling.

Two recent papers examine the use of LLMs to predict both individual and societal policy preferences, as an 'augmentation' to current democratic input methods. Gudiño-Rosero and colleagues (2024) argue that augmented democracy, in this context, "is based on the construction of digital twins, which in the case of this paper, were constructed using a minimalistic representation of each agent (a relatively short vector or demographic characteristics and pairwise preferences)." The idea is to create a simulated version of an individual citizen that can "represent" that person's policy preferences in some arenas of democratic input, as with AI assistants in digital commercial transactions. The authors continue that augmented democracy is:

> …different from the creation of AI politicians, as it does not involve the creation of a single AI representative designed to 'listen to everyone,' but an ensemble of AI agents, each controlled by its own human: citizens can create individual profiles that can be personalized according to their own characteristics, preferences and habits, and these autonomous agents can potentially vote on their behalf.[43]

This work stops far short of the creation of actual "digital twins" for political representation purposes. However, one can imagine personalized LLMs with long-term memory capable of a high-quality simulation (that is, after an extended conversation with the subject and appropriate background information).

---

[41] With lots of caveats.
[42] https://arxiv.org/abs/2412.04498
[43] https://arxiv.org/abs/2405.03452

Another recent contribution (Majumdar et al., 2024) considers LLMs' capacity to accurately simulate elections.[44] "Generative AI Voting: Fair Collective Choice is Resilient to LLM biases and Inconsistencies" shows that, under specific circumstances, LLMs "prove to be a win-win: fairer voting outcomes for humans with fairer AI representation." For simple ballots (say, two national candidates for president), "consistency of large language models becomes significant under fair voting aggregation methods that promote a proportional representation of voters' preferences." This paper indicates that while LLMs don't yet approximate 'digital twins,' when they are applied to simple ballots at the aggregate level, they simulate consistent results.

These findings suggest three important takeaways. One, a sufficiently high-fidelity simulation would represent (at least) indirect democratic input. If your 'digital twin' accurately represents your interests in 99% of cases, it seems reasonable to make them your proxy, in a variety of political forums, if/when you lack the time or contextual knowledge to represent yourself. This circumstance would increase the amount of citizen feedback expressed directly to the government, to such a degree that the role of 'elected representative' may come to suggest oversight of the AGI system rather representation of human constituents. Would elected representatives become obsolete or undesirable for these tasks? Would they be willing (or even able) to qualitatively represent the interests of their electorate?

Secondly, an AGI system should be able to effectively 'set the agenda' for policy deliberations, further obfuscating the need for elected representatives in a legislative or parliamentary system. The AGI system could generate proposals, create a ranking system, and solicit and receive feedback from digital twins and humans as needed. Finally, depending on the technological trajectory of AGI, a high volume of experimentation, ideally at the local level or alongside other current systems, would help illustrate the best socio-technical arrangement of these new democratic capabilities.

**Concerns**
In addition to their potential for harnessing democratic input, LLM systems are already being deployed to assist, or augment, democratic participation, raising worries about unintended consequences. Aoki et al. (2024) discuss the challenges faced by LLMs in the realms of bias, transparency, and accountability. Majumdar and colleagues (2024) echo their concern:

> For instance, who shall determine the input training data of AI representatives? Should the training data involve only self-determined personal information of voters, or shall these be augmented with more universal knowledge and experts' opinions? How to protect the privacy and autonomy of voters when training such AI representatives? Will citizens retain power to control AI representatives that reflect their values and beliefs, while remaining accountable?

---

[44] https://arxiv.org/abs/2406.11871

In addition to the above questions (which also apply to AGI in general), the integration of AGI into the project of democratic input presents at least three major concerns: alignment, accuracy, and dignity. First, how can we be sure that the AGI does what we want it to do? The enmeshment of unaligned AGI systems with democratic input processes makes them manipulable, distorting the democratic feedback process. Second, how confident are we with the accuracy of digital twins and societal-level simulations? Are all relevant factors indeed available, measurable, and measured? Finally, what impact might AGI involvement have on human dignity? By delegating our democratic input to a digital proxy, do we also throw away our chance to engage meaningfully in our joint democratic endeavor?

## VI. Securing the Narrow Corridor in the Age of AGI

As we have explored throughout this essay, the advent of artificial general intelligence (AGI) represents a potential step-change in the capabilities of both state and societal actors. Just as past technological revolutions (e.g., the printing press and industrialization) reshaped the institutions of governance, AGI is likely to disrupt the delicate balance of power that sustains free societies.

Historically, liberal societies have maintained a precarious equilibrium between state capacity and individual liberty, with constitutional constraints, checks and balances, and the rule of law serving to keep both despotism and anarchy at bay. However, this equilibrium has never been static. Technological and social change have forced repeated renegotiations of the social contract, from the rise of mass politics in the industrializing West to the welfare state reforms of the early 20$^{th}$ century.

AGI represents a similar inflection point. AGI's transformative capabilities risk destabilizing the narrow corridor, in two directions. On one hand, it could dramatically enhance the administrative capacity of the state, enabling more effective provision of public goods and services. Governments that successfully harness AGI could become radically more efficient, responsive, and data-driven in their decision-making. Moreover, AGI systems could potentially help to address long-standing challenges in public administration, such as coordination failures and principal-agent problems.

However, this same enhancement of state capacity also risks a slide towards authoritarianism, the 'despotic Leviathan,' if not carefully constrained. An AGI-empowered state could wield unprecedented surveillance and control over its citizens, stifling dissent and entrenching existing power structures. The increasing automation of administrative decision-making could also erode human agency and democratic accountability, as bureaucracies evolve towards more centralized, 'system-level' architectures.

On the other hand, if AGI diffuses more rapidly among individuals and civil society groups than governments, it could instead weaken the legitimacy and capacity of the state relative to non-state actors. In this scenario, the 'absent Leviathan,' the risk is not despotism but a hollowing out of the governability and social cohesion that liberal democracies depend upon. Malicious actors could also exploit widely accessible AGI to undermine elections, manipulate public opinion, or coordinate insurgencies, further eroding the stability of democratic institutions.

The integration of AGI into public administration further complicates this balance by fundamentally altering how governments function. Governments are not monolithic entities, but rather complex systems of agencies and bureaucracies tasked with diverse and often conflicting objectives. AGI will likely accelerate the evolution of these systems, pushing them toward system-level bureaucracies where decision-making and execution increasingly rely on automated systems. For example, AGI systems acting as artificial bureaucrats could automate complex, high-discretion tasks that traditionally required human judgment, promising significant gains in efficiency, effectiveness, and scalability and transforming how governments operate. However, their widespread deployment raises critical questions about transparency, accountability, and responsiveness—values essential to liberal democratic governance.

While AGI offers transformative benefits, it also presents risks to the administrative values that underpin effective governance in free societies. AGI systems could dramatically improve the efficiency and effectiveness of government tasks, reducing costs and increasing output quality. However, its impact on equity is less certain, as AGI systems may inadvertently replicate or exacerbate societal biases embedded in their training data. Moreover, transparency and accountability could suffer as AGI systems make decisions based on processes that are opaque to both administrators and the public. Mechanistic interpretability and explainable AI research provide some tools for addressing these challenges, but aligning AGI systems with democratic principles will remain an ongoing and complex task.

The ethical implications of AGI's integration into governance are equally significant. Delegating value-laden decisions to AGI raises concerns about the loss of moral accountability in public administration. For example, while AGI agents may excel at optimizing policies for efficiency, they may lack the ethical nuance required to address competing societal values. This disconnect between computational optimization and human morality risks eroding public trust, particularly if AGI systems prioritize narrow objectives at the expense of fairness and inclusivity. Ensuring that AGI agents reflect societal priorities and democratic values is therefore critical to preserving the integrity of governance systems.

To address these challenges and secure the narrow corridor, a comprehensive strategy must include technological safeguards, institutional adaptations, and ethical considerations. On the technological front, privacy-enhancing technologies (PETs) can play a vital role in counterbalancing the surveillance capabilities of AGI. Advanced PETs could enable individuals to maintain autonomy and privacy in the face of increasingly pervasive state monitoring. Simultaneously, investments in explainable AI and mechanistic interpretability are essential to ensuring that AGI systems operate transparently and remain accountable for their decisions. These technologies can help bridge the gap between AGI's computational processes and human oversight, fostering trust in their deployment.

Institutional adaptations are equally important. Governments must embrace hybrid AI-human governance structures that combine AGI's computational power with the nuanced judgment and accountability that human administrators provide. Bureaucratic models must also evolve to maintain flexibility and adaptability, allowing agencies to leverage AGI's capabilities while retaining oversight mechanisms that align with democratic values. Such adaptations will be crucial in preserving the legitimacy of governance systems as they confront the rapid pace of technological change. At the same time, policymakers will need to adapt regulatory and legal frameworks to account for the increasing use of AGI by non-state actors, whether in setting and enforcing standards or modernizing antitrust paradigms to address market concentration risks.

Reinforcing democratic processes is another critical pillar of securing the narrow corridor. AGI systems offer unique opportunities to enhance participatory governance, such as enabling large-scale deliberative platforms, real-time citizen feedback systems, and representative digital twins. These tools could revitalize democratic engagement and strengthen the feedback loop between citizens and their representatives. However, these systems must be designed with robust safeguards to prevent misuse and ensure that they genuinely enhance, rather than undermine, democratic accountability. Establishing clear frameworks for democratic oversight of AGI systems will also be essential, ensuring that artificial bureaucrats remain accountable to elected officials and the public.

Perhaps most importantly, securing the narrow corridor in an age of AGI will require an epistemic shift in how we approach the governance of emerging technologies. Rather than passively reacting to technological disruptions, policymakers and publics alike must cultivate a greater capacity for anticipatory governance—proactively imagining and stress-testing institutional paradigms in expectation of AGI's transformative potential. Intellectual frameworks like scenario planning, threat modeling, and forecasting should be deployed, in a serious exploration of failure modes and policy options for their mitigation. Governance experimentation across scales—from local sandboxes to international norm-setting—will also be needed to surface gaps in existing governance frameworks and prototype more AGI-robust alternatives.

Achieving such a shift will not be easy. It requires a far more dynamic and adaptive approach to policymaking than most governments are accustomed to. It also demands greater collaboration between AI researchers, social scientists, and policymakers in exploring the societal implications of AGI. And it necessitates reckoning honestly with the ways in which existing political

institutions—from the administrative state to representative democracy itself—may be inadequate to the challenges posed by artificial superintelligence.

Meeting these challenges is nonetheless essential if we are to ensure that the immense power of AGI is channeled towards expanding rather than constraining human freedom and flourishing. The great political question of the 21$^{st}$ century may well be whether liberal democracy can reform itself in time to reap the rewards and manage the risks of artificial general intelligence.

History suggests that, with sufficient foresight and resolve, free societies are capable of extraordinary institutional innovation in moments of technological upheaval. The task before us is to muster that resolve once again—to reimagine the narrow corridor for an age of AGI—and in so doing, secure the possibility of a brighter future for all. Only by rising to this challenge can we ensure that the tremendous power of AGI remains firmly in service of our deepest democratic values: individual liberty, popular sovereignty, and human dignity for all.

Heldring, L., Robinson, J., & Vollmer, S. (2022). *The Economic Effects of the English Parliamentary Enclosures* (No. w29772; p. w29772). National Bureau of Economic Research. https://doi.org/10.3386/w29772

Ho, A., Besiroglu, T., Erdil, E., Owen, D., Rahman, R., Guo, Z. C., Atkinson, D., Thompson, N., & Sevilla, J. (2024). *Algorithmic progress in language models* (No. arXiv:2403.05812). arXiv. https://doi.org/10.48550/arXiv.2403.05812

Humlum, A., & Vestergaard, E. (2024). The Adoption of ChatGPT. *SSRN Electronic Journal*. https://doi.org/10.2139/ssrn.4827166

Johnson, N. D., & Koyama, M. (2019). *Persecution and Toleration: The Long Road to Religious Freedom* (1st ed.). Cambridge University Press. https://doi.org/10.1017/9781108348102

Kiela et al. (2023). *Test scores of the AI relative to human performance* [Dataset]. https://ourworldindata.org/grapher/test-scores-ai-capabilities-relative-human-performance

Korinek, A. (2023). *Language Models and Cognitive Automation for Economic Research* (No. w30957; p. w30957). National Bureau of Economic Research. https://doi.org/10.3386/w30957

Korinek, A., & Juelfs, M. (2022). *Preparing for the (Non-Existent?) Future of Work* (No. w30172; p. w30172). National Bureau of Economic Research. https://doi.org/10.3386/w30172